\documentclass[twocolumn,preprintnumbers,amsmath,amssymb,prb]{revtex4}

\usepackage{graphicx}
\usepackage{dcolumn}
\usepackage{bm}
\usepackage{color}

\begin{document}

\preprint{}

\title{Elastic softness of low-symmetry frustrated $A$Ti$_2$O$_5$ ($A$ = Co, Fe)}

\author{Tadataka Watanabe$^1$}
\thanks{Electronic address: watanabe.tadataka@nihon-u.ac.jp}
\author{Kazuya Takayanagi$^1$}
\author{Ray Nishimura$^1$}
\author{Yoshiaki Hara$^2$}
\author{Dharmalingam Prabhakaran$^3$}
\author{Roger D. Johnson$^4$}
\author{Stephen J. Blundell$^3$}
\affiliation{$^1$Department of Physics, College of Science and Technology, Nihon University, Chiyoda, Tokyo 101-8308, Japan}
\affiliation{$^2$National Institute of Technology, Ibaraki College, Hitachinaka 312-8508, Japan}
\affiliation{$^3$Department of Physics, Clarendon Laboratory, Oxford University, Parks Road, Oxford OX1 3PU, United Kingdom}
\affiliation{$^4$Department of Physics and Astronomy and London Centre for Nanotechnology, University College London, London WC1E 6BT, United Kingdom}
\date{\today}

\begin{abstract}
Orthorhombic pseudobrookites CoTi$_2$O$_5$ and FeTi$_2$O$_5$ have a low-symmetry crystal structure comprising magnetic Co$^{2+}$/Fe$^{2+}$ ions and nonmagnetic Ti$^{4+}$ ions, where the orbital-nondegenerate Co$^{2+}$/Fe$^{2+}$ ions form one-dimensional chains running along the orthorhombic $a$ axis. These compounds undergo an antiferromagnetic phase transition at $T_N \sim$ 26 K for CoTi$_2$O$_5$ and $T_N \sim$ 40 K for FeTi$_2$O$_5$. Ultrasound velocity measurements on single crystals of CoTi$_2$O$_5$ and FeTi$_2$O$_5$ reveal that CoTi$_2$O$_5$ exhibits unusual elastic softness above $T_N$ in the symmetry-lowering elastic mode of $ac$-plane shear elastic modulus, inconsistent with the structural symmetry breaking caused by antiferromagnetic ordering at $T_N$. This suggests the presence of two distinct types of magnetostructural fluctuations above $T_N$ that should be a precursor to the symmetry-lowering lattice distortion at $T_N$. In contrast, FeTi$_2$O$_5$ exhibits either negligible or smaller elastic softness, indicating weaker spin-lattice coupling. These findings highlight CoTi$_2$O$_5$ and FeTi$_2$O$_5$ as unique spin-latticed-coupled frustrated systems with low crystal symmetry, where, while the exchange interactions are quasi-one-dimensional in nature, the frustration is released by further lowering the crystal symmetry through three-dimensional spin-lattice coupling, which is stronger in CoTi$_2$O$_5$ than in FeTi$_2$O$_5$.
\end{abstract}

\maketitle

\section{Introduction}

Frustrated magnets provide fertile ground for exploring novel correlated phenomena emerging from competing magnetic interactions [\cite{Lacroix}]. In a frustrated system with spin-lattice coupling, the spin degeneracy can be lifted by symmetry-lowering lattice distortion [\cite{Yamashita,Tchernyshyov}]. This effect is called the spin Jahn--Teller (spin-JT) effect because its mechanism is analogous to the Jahn--Teller effect in orbital-degenerate systems where the spontaneous lattice distortion reduces the crystal symmetry to lift the orbital degeneracy [\cite{Gehring}]. The spin-JT effect was initially proposed for cubic pyrochlore antiferromagnets where spins form a lattice of corner-sharing tetrahedra [\cite{Yamashita,Tchernyshyov}]. For real pyrochlore antiferromagnets, the spin-JT mechanism has been applied to explain the magnetostructural transitions of some cubic spinels, where a cubic-to-tetragonal lattice distortion releases the frustration [\cite{Lee,Ortega-San-Martin,Chung}].

Recently, orthorhombic pseudobrookites $A$Ti$_2$O$_5$ ($A$ = Co, Fe), having much lower symmetry than the prototypical spin-JT system of cubic spinels, have been proposed as candidate spin-JT systems [\cite{Kirschner,Lang}]. These pseudobrookites comprise magnetic Co$^{2+}$ (3$d^7$, $S$ = 3/2) or Fe$^{2+}$ (3$d^6$, $S$ = 2) ions and nonmagnetic Ti$^{4+}$ ions. In addition, the magnetic $A^{2+}$ ions occupying the equivalent crystal sites form one-dimensional (1D) chains running along the orthorhombic $a$ axis [Fig. 1]. In $A$Ti$_2$O$_5$, the magnetic $A^{2+}$ ions with $C_{2v}$ site symmetry have no orbital degeneracy [Fig. 1(*)], and the interchain exchange interactions are considered to be frustrated [\cite{Kirschner,Lang}]. For FeTi$_2$O$_5$, the absence of orbital degeneracy was confirmed by Wannier function projection of nonmagnetic generalized-gradient-approximation calculations for the experimental crystal structure [Fig. 1(*)] [\cite{Xu}].

$A$Ti$_2$O$_5$ undergoes an antiferromagnetic (AF) transition at $T_N \sim$ 26 K for CoTi$_2$O$_5$ and $T_N \sim$ 40 K for FeTi$_2$O$_5$ [\cite{Kirschner,Lang,Xu2,Xu}]. For CoTi$_2$O$_5$, the long-range AF order with the propagation vector {\bf q} = ($\pm\frac{1}{2},\frac{1}{2},$0) has been identified in neutron powder diffraction and muon-spin rotation experiments [\cite{Kirschner}]. The emergence of this AF order requires a symmetry-lowering lattice distortion for the orthorhombic CoTi$_2$O$_5$. Thus, this orbital-nondegenerate frustrated antiferromagnet is suggested to be a spin-JT magnet. For FeTi$_2$O$_5$, by combining muon spin rotation and x-ray diffraction experiments with density functional theory calculations, it has been demonstrated that the crystal and AF structures are the same as those of CoTi$_2$O$_5$ [\cite{Lang}]. This suggests that not only CoTi$_2$O$_5$ but also FeTi$_2$O$_5$ are spin-JT magnets. Although the symmetry-lowering lattice distortion in the AF state has not yet been experimentally resolved in either CoTi$_2$O$_5$ or FeTi$_2$O$_5$, recent resonant elastic x-ray scattering experiments in CoTi$_2$O$_5$ and thermal expansion and magnetostriction measurements in FeTi$_2$O$_5$ revealed the presence of magnetoelastic coupling, which supports the spin-JT scenario in $A$Ti$_2$O$_5$ [\cite{Behr,Xu}].

\begin{figure}[t]
\begin{center}
\includegraphics[scale=0.45]{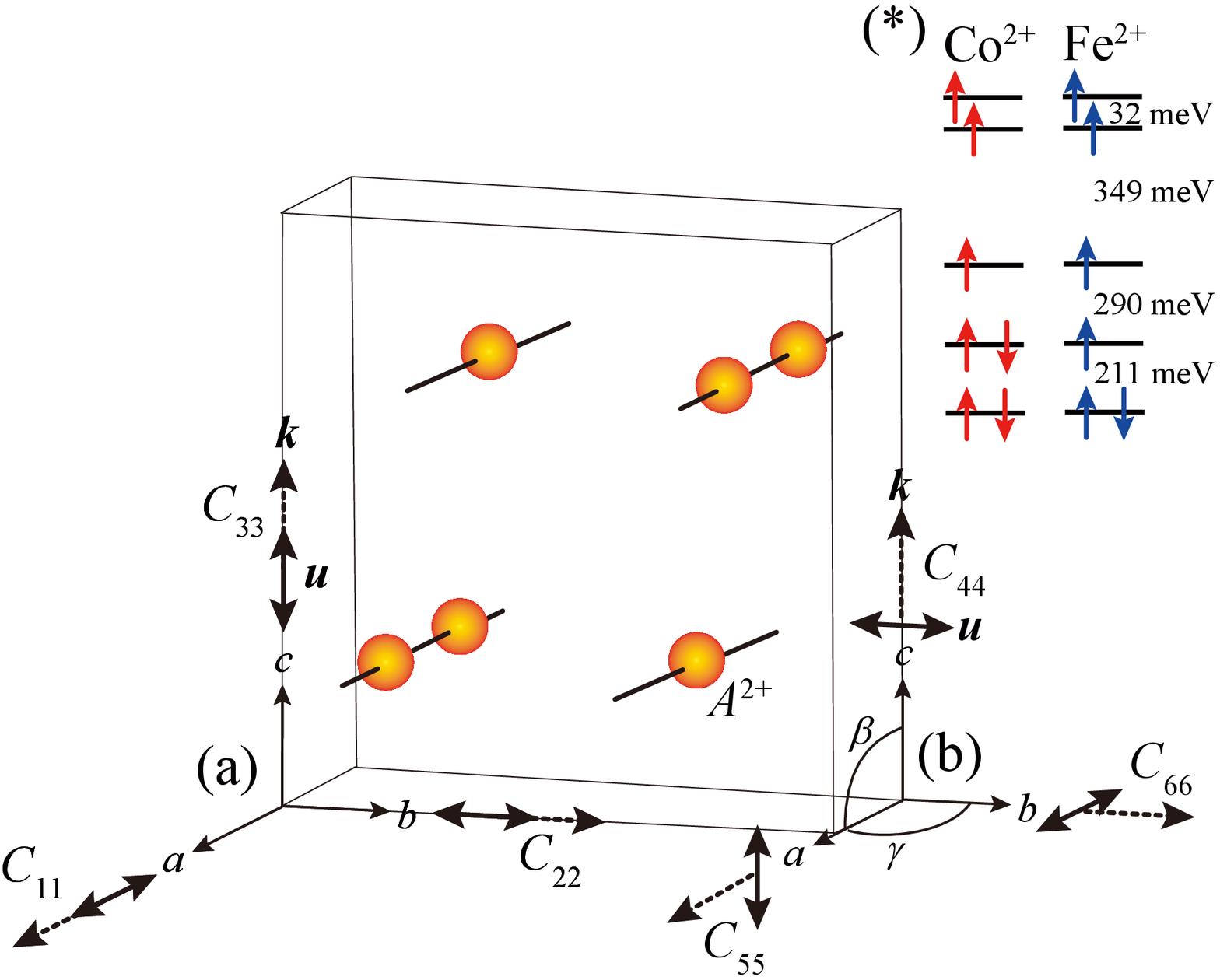}
\caption{\label{fig:fig1} (Color online) Magnetic $A^{2+}$ sites in $A$Ti$_2$O$_5$ ($A$ = Co, Fe) within the orthorhombic crystal unit cell, forming 1D $a$-axis chains (the solid lines). (*) Schematic energy levels of Co$^{2+}$ 3$d^7$ and Fe$^{2+}$ 3$d^6$ electrons with $C_{2v}$ site symmetry in $A$Ti$_2$O$_5$. The values in (*) represent the crystal-field splitting energies in FeTi$_2$O$_5$ as reported in Ref. [\cite{Xu}]. With the crystal axes (a) and (b), the propagation vector {\bf k} and polarization vector {\bf u} of the sound waves for (a) compressive elastic moduli $C_{11}$, $C_{22}$, and $C_{33}$, and (b) shear elastic moduli $C_{44}$, $C_{55}$, and $C_{66}$ are indicated. The angles $\beta$ and $\gamma$ between the crystal axes (b) correspond to the shear angles that are respectively tilted by the sound waves for $C_{55}$ ({\bf k} $\parallel$ {\bf a}, {\bf u} $\parallel$ {\bf c}) and $C_{66}$ ({\bf k} $\parallel$ {\bf b}, {\bf u} $\parallel$ {\bf a}).}
\end{center}
\end{figure}

In this paper, we present ultrasound velocity measurements of the low-symmetry orthorhombic $A$Ti$_2$O$_5$ ($A$ = Co, Fe), from which we determine the elastic moduli of these compounds. The sound velocity or elastic modulus is a useful probe with which to extract symmetry-resolved thermodynamic information from a crystal [\cite{Luthi}]. Furthermore, as the ultrasound velocity can be measured with a high precision of approximately parts per million, its measurements can sensitively probe elastic anomalies driven by phase transitions, fluctuations, and excitations [\cite{Luthi}]. In magnets, the modified sound dispersions caused by magnetoelastic coupling allow the extraction of detailed information on the interplay of the lattice, spin, and orbital degrees of freedom [\cite{Luthi,Bhattacharjee,Watanabe1,Watanabe2,Watanabe3,Nii,Watanabe4,Watanabe5,Watanabe6,Watanabe7,Watanabe8,Watanabe9,Kino,Kataoka,Hazama,Kindler,Ramshaw,Saint-Paul,Poirier,Schwenk}].

In the spin-JT system, spin-lattice coupling induces a structural transition that lowers the crystal symmetry. For instance, the representative spin-JT system of chromite spinel $A$Cr$_2$O$_4$ ($A$ = Mg, Zn) has an AF transition at $T_N \sim$ 13 K coinciding with the cubic-to-tetragonal lattice distortion [\cite{Lee,Ortega-San-Martin}]. For $A$Cr$_2$O$_4$, the ultrasound velocity measurements in the single crystal revealed that the temperature ($T$) dependence of the tetragonal shear modulus ($C_{11}-C_{12}$)/2 exhibits Curie-type ($\sim-1/T$-type) softening upon cooling in the cubic paramagnetic phase ($T>T_N$), which is a precursor to the cubic-to-tetragonal lattice distortion at $T_N$ [\cite{Watanabe3}]. This elastic anomaly above $T_N$ indicates the presence of spin-lattice-coupled fluctuations in the paramagnetic phase of $A$Cr$_2$O$_4$, which is referred to as the dynamical spin-JT effect. In the present study of the orthorhombic pseudobrookites CoTi$_2$O$_5$ and FeTi$_2$O$_5$, we find the presence of unusual elastic softness in CoTi$_2$O$_5$ in the paramagnetic phase above $T_N$, suggesting the emergence of dynamical spin-lattice-coupled state.

\begin{table}[b]
\caption{\label{tab:table1} Elastic moduli for $A$Ti$_2$O$_5$ ($A$ = Co, Fe) with an orthorhombic crystal structure, and the corresponding sound mode (propagation vector {\bf k} and polarization vector {\bf u}) and irreducible representation (irrep).}
\begin{ruledtabular}
\begin{tabular}{ccc}
Elastic modulus & Sound mode ({\bf k} and {\bf u}) & Irrep\\
\hline
$C_{11}$ & Longitudinal wave ({\bf k} $\parallel$ {\bf u} $\parallel$ {\bf a}) & A$_g$\\
$C_{22}$ & Longitudinal wave ({\bf k} $\parallel$ {\bf u} $\parallel$ {\bf b}) & A$_g$\\
$C_{33}$ & Longitudinal wave ({\bf k} $\parallel$ {\bf u} $\parallel$ {\bf c}) & A$_g$\\
\hline
$C_{44}$ & Transverse wave ({\bf k} $\parallel$ {\bf c}, {\bf u} $\parallel$ {\bf b}) & B$_{3g}$\\
$C_{55}$ & Transverse wave ({\bf k} $\parallel$ {\bf a}, {\bf u} $\parallel$ {\bf c}) & B$_{2g}$\\
$C_{66}$ & Transverse wave ({\bf k} $\parallel$ {\bf b}, {\bf u} $\parallel$ {\bf a}) & B$_{1g}$\\
\end{tabular}
\end{ruledtabular}
\end{table}

\section{Experimental}

Single crystals of CoTi$_2$O$_5$ with $T_N\sim$ 26 K and FeTi$_2$O$_5$ with $T_N\sim$ 40 K were grown adopting the floating-zone method [\cite{Kirschner,Lang}]. The ultrasound velocities were measured adopting the phase-comparison technique with longitudinal and transverse sound waves at a frequency of 30 MHz, where the ultrasound velocity or elastic modulus can be measured with a high precision of approximately parts per million. The ultrasound waves were generated and detected by LiNbO$_3$ transducers with a fundamental frequency of $f$ = 30 MHz, which were attached to parallel mirror surfaces of the crystal, oriented perpendicular to the orthorhombic $a$, $b$, and $c$ axes. The variation in ultrasound attenuation was monitored by observing the amplitude of the first transmitted echo. In the present experiments, ultrasound signals at higher-harmonic frequencies were not detected. Measurements were taken to determine the symmetrically independent elastic moduli of the orthorhombic crystal, specifically the compressive elastic moduli $C_{11}$, $C_{22}$, and $C_{33}$, and the shear elastic moduli $C_{44}$, $C_{55}$, and $C_{66}$ (see Table I). In Fig. 1, the propagation vector {\bf k} and polarization vector {\bf u} of the sound waves for the respective elastic moduli are indicated along with the magnetic $A^{2+}$ sites of $A$Ti$_2$O$_5$ ($A$ = Co, Fe) in the orthorhombic crystal unit cell, which form the 1D $a$-axis chains. As indicated in Fig. 1(a), the longitudinal sound wave corresponding to the compressive elastic modulus $C_{11}$ propagates along the magnetic $A^{2+}$ chains ({\bf k} $\parallel$ {\bf a}), whereas the longitudinal waves corresponding to the compressive moduli $C_{22}$ and $C_{33}$ propagate orthogonal to the $A^{2+}$ chains ({\bf k} $\perp$ {\bf a}). Likewise, in Fig. 1(b), the transverse sound wave corresponding to the shear elastic modulus $C_{55}$ propagates along the $A^{2+}$ chains ({\bf k} $\parallel$ {\bf a}), whereas the transverse waves corresponding to the shear $C_{44}$ and $C_{66}$ propagate orthogonal to the $A^{2+}$ chains ({\bf k} $\perp$ {\bf a}). The sound velocities of CoTi$_2$O$_5$ (FeTi$_2$O$_5$) measured at a room temperature of 300 K are 8820 m/s (8600 m/s) for $C_{11}$, 8750 m/s (6380 m/s) for $C_{22}$, 8630 m/s (7620 m/s) for $C_{33}$, 3920 m/s (3600 m/s) for $C_{44}$, 4300 m/s (4090 m/s) for $C_{55}$, and 4040 m/s (3660 m/s) for $C_{66}$.

\section{Results}
\subsection{CoTi$_2$O$_5$}

\begin{figure}[t]
\begin{center}
\includegraphics[scale=0.6]{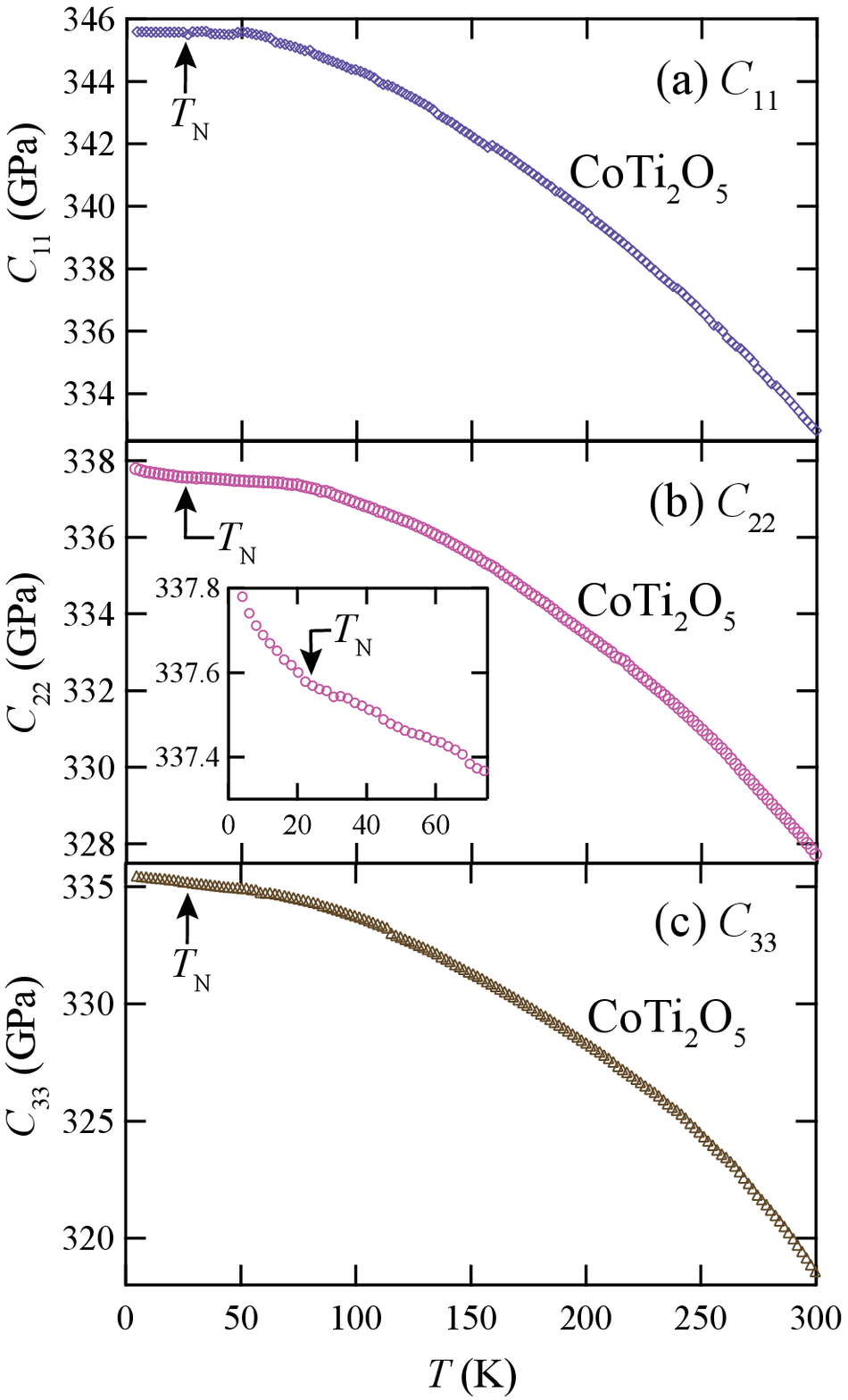}
\caption{\label{fig:fig2} (Color online) Compressive elastic moduli of CoTi$_2$O$_5$ as functions of $T$: (a) $C_{11}(T)$, (b) $C_{22}(T)$, and (c) $C_{33}(T)$. The inset in (b) shows the expanded view of $C_{22}(T)$ below 75 K. The labeled arrows indicate $T_N \sim$ 26 K of CoTi$_2$O$_5$.}
\end{center}
\end{figure}

\begin{figure}[t]
\begin{center}
\includegraphics[scale=0.6]{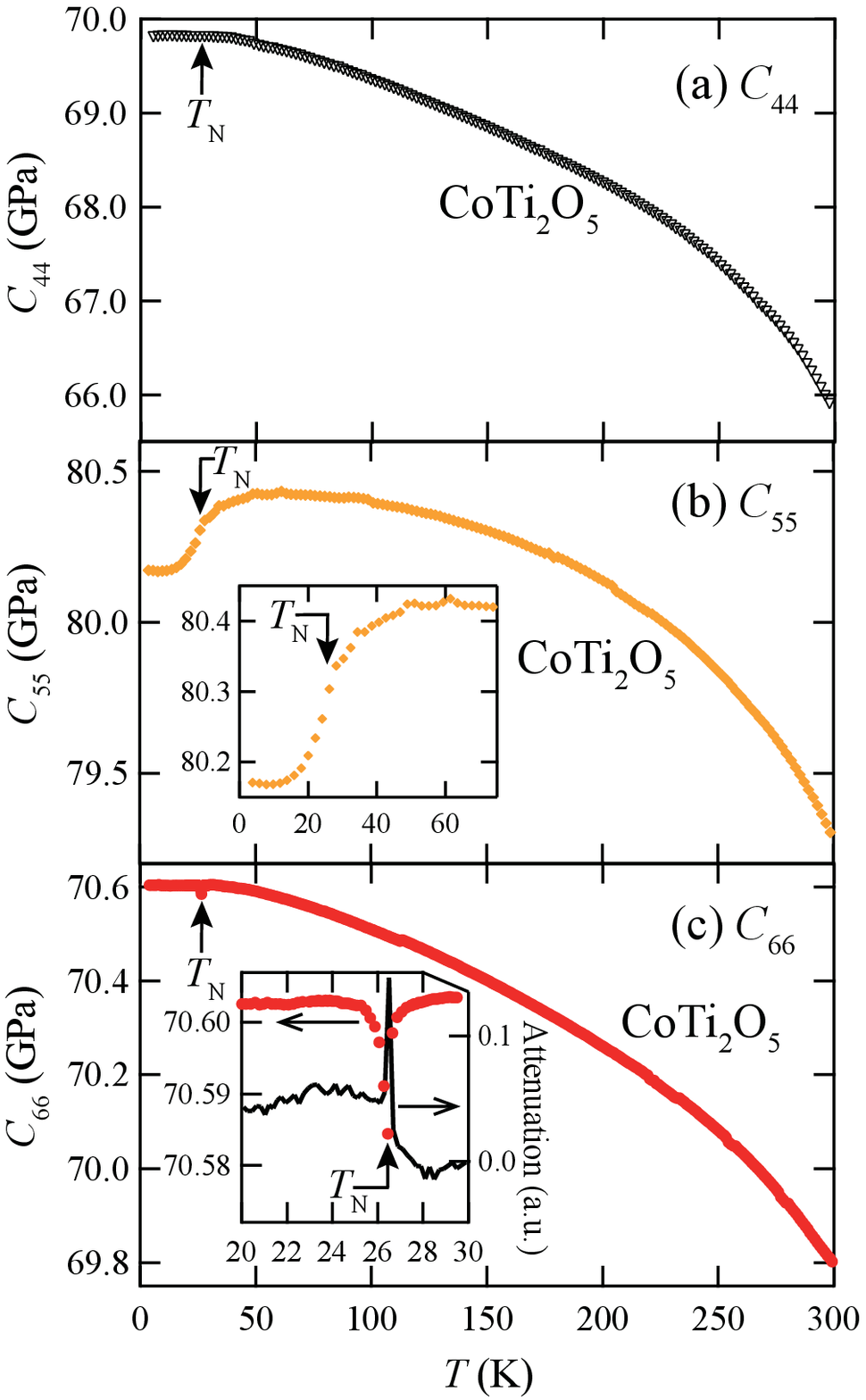}
\caption{\label{fig:fig3} (Color online) Shear elastic moduli of CoTi$_2$O$_5$ as functions of $T$: (a) $C_{44}(T)$, (b) $C_{55}(T)$, and (c) $C_{66}(T)$. The inset in (b) is the expanded view of $C_{55}(T)$ below 75 K. The inset in (c) displays the expanded view of $C_{66}(T)$ [markers] along with the relative $T$ dependence of the ultrasound attenuation for $C_{66}$ normalized to its value at 30 K [solid line], in the range 20 K $\leq T \leq$ 30 K. The labeled arrows indicate $T_N \sim$ 26 K of CoTi$_2$O$_5$.}
\end{center}
\end{figure}

Figure 2(a)--(c) respectively present the temperature ($T$) dependence of the compressive elastic moduli $C_{11}(T)$, $C_{22}(T)$, and $C_{33}(T)$ in CoTi$_2$O$_5$. $C_{11}(T)$ [Fig. 2(a)] and $C_{33}(T)$ [Fig. 2(c)] exhibit monotonic hardening upon cooling from 300 to 2 K, as is usually observed in solids [\cite{Varshni}]. In addition, $C_{22}(T)$ [Fig. 2(b)] exhibits ordinary monotonic hardening upon cooling from 300 to $T_N\sim$ 26 K and a small slope change of hardening at $T_N\sim$ 26 K.

Figure 3(a)--(c) respectively depict the $T$ dependence of the shear elastic moduli $C_{44}(T)$, $C_{55}(T)$, and $C_{66}(T)$ in CoTi$_2$O$_5$. In these plots, all the elastic moduli exhibit hardening upon cooling from 300 to $\sim$60 K, but below $\sim$60 K, only $C_{55}(T)$ [Fig. 3(b)] exhibits Curie-type ($\sim-1/T$-type) softening upon cooling to $T_N\sim$ 26 K. Furthermore, $C_{55}(T)$ [Fig. 3(b)] and $C_{66}(T)$ [Fig. 3(c)] exhibit a discontinuous anomaly at $T_N\sim$ 26 K, whereas there is no anomaly in $C_{44}(T)$ [Fig. 3(a)]. Near and below $T_N$, the ultrasound signals for $C_{55}(T)$ and $C_{66}(T)$ are attenuated but remain strong enough to accurately measure the $T$ dependence of the absolute values of the sound velocities. As highlighted in the inset of Fig. 3(c), the $T$ dependence of the ultrasound attenuation for $C_{66}$ exhibits a sharp peak at $T_N\sim$ 26 K, coinciding with the dip anomaly observed in $C_{66}(T)$.

In $C_{55}(T)$ [Fig. 3(b)], the Curie-type softening ceases below $T_N$, but the elasticity does not recover in the AF state below $T_N$, where the detected ultrasound signal is attenuated. In the magnetically ordered state, magnetostriction can induce domain-wall stress, resulting in a loss of elasticity [\cite{Bozorth}]. Thus, the softened elasticity observed in $C_{55}(T)$ below $T_N$ can be attributed to stress effects on the magnetic domain walls.

\subsection{FeTi$_2$O$_5$}

Figure 4(a)--(c) present the $T$ dependence of the compressive elastic moduli $C_{11}(T)$, $C_{22}(T)$, and $C_{33}(T)$, respectively, in FeTi$_2$O$_5$. $C_{11}(T)$ [Fig. 4(a)] and $C_{22}(T)$ [Fig. 4(b)] exhibit monotonic hardening upon cooling from 300 to 2 K, as is usually observed in solids [\cite{Varshni}]. Moreover, $C_{33}(T)$ [Fig. 4(c)] exhibits ordinary monotonic hardening upon cooling from 300 to $T_N\sim$ 40 K and a small slope change of the hardening at $T_N\sim$ 40 K.

Figure 5(a)--(c) depict the $T$ dependence of the shear elastic moduli $C_{44}(T)$, $C_{55}(T)$, and $C_{66}(T)$, respectively, in FeTi$_2$O$_5$. $C_{66}(T)$ [Fig. 5(c)] exhibits monotonic hardening upon cooling from 300 to 2 K. Moreover, $C_{44}(T)$ [Fig. 5(a)] and $C_{55}(T)$ [Fig. 5(b)] exhibit monotonic hardening upon cooling from 300 K to $T_N\sim$ 40 K and a small slope change of the hardening at $T_N\sim$ 40 K.

\begin{figure}[t]
\begin{center}
\includegraphics[scale=0.6]{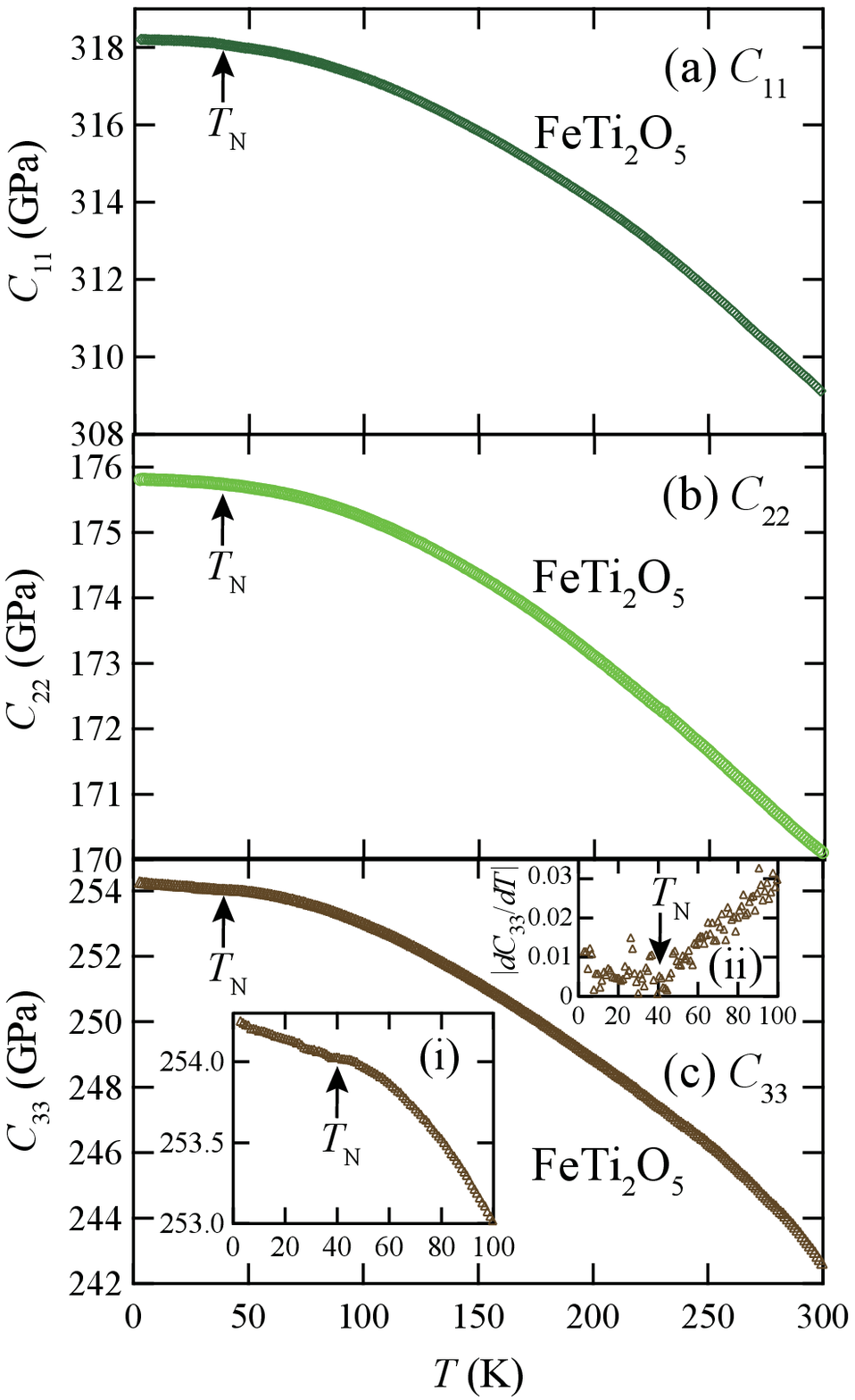}
\caption{\label{fig:fig4} (Color online) Compressive elastic moduli of FeTi$_2$O$_5$ as functions of $T$: (a) $C_{11}(T)$, (b) $C_{22}(T)$, and (c) $C_{33}(T)$. The insets (i) and (ii) in (c) respectively show the expanded views of $C_{33}(T)$ and $|dC_{33}(T)/dT|$ below 100 K. The labeled arrows indicate $T_N \sim$ 40 K of FeTi$_2$O$_5$.}
\end{center}
\end{figure}

\begin{figure}[t]
\begin{center}
\includegraphics[scale=0.6]{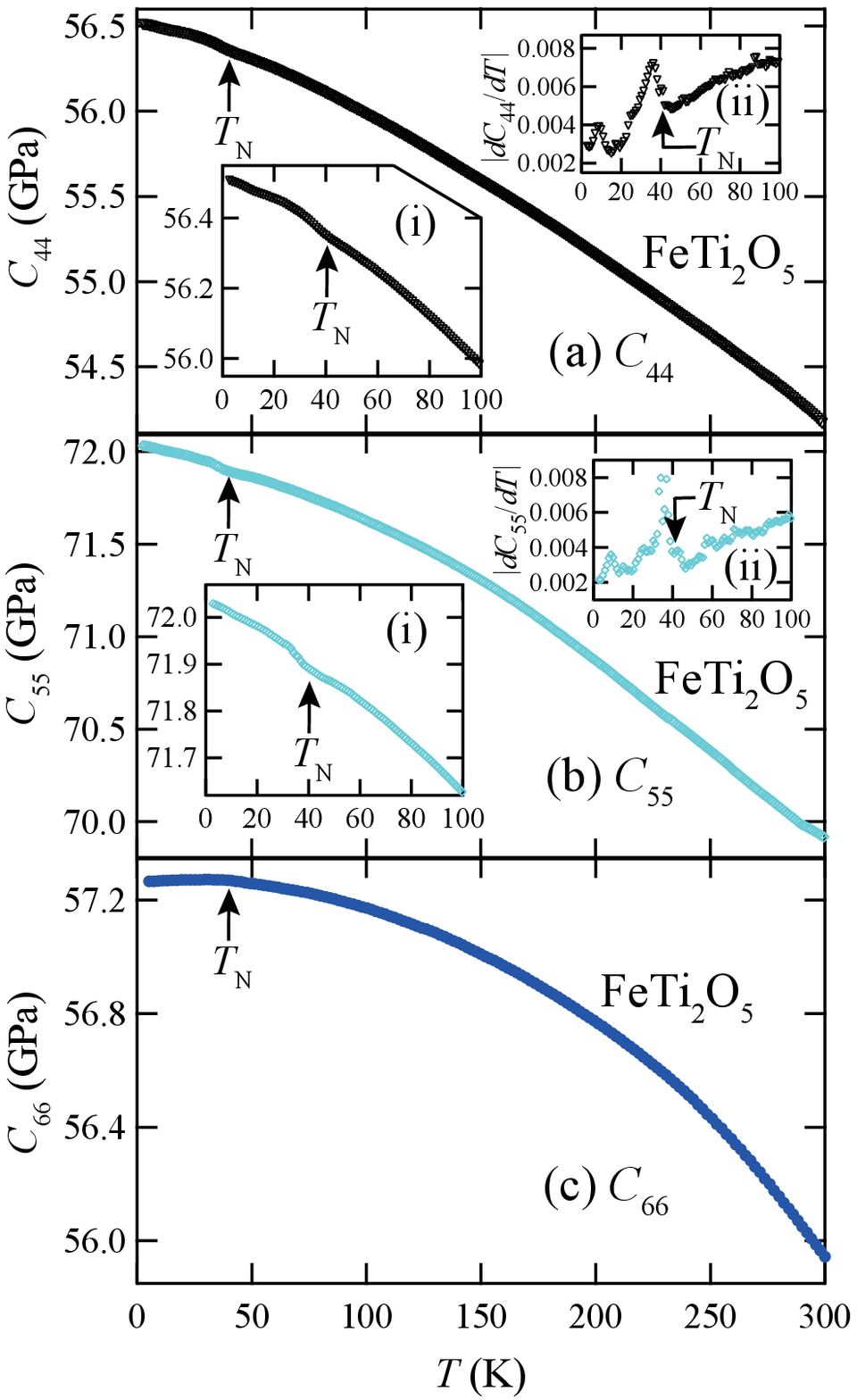}
\caption{\label{fig:fig5} (Color online) Shear elastic moduli of FeTi$_2$O$_5$ as functions of $T$: (a) $C_{44}(T)$, (b) $C_{55}(T)$, and (c) $C_{66}(T)$. The insets (i) and (ii) in (a) and (b) respectively show the expanded views of $C_{55}(T)$, $|dC_{55}(T)/dT|$, $C_{66}(T)$, and $|dC_{66}(T)/dT|$ below 100 K. The labeled arrows indicate $T_N \sim$ 40 K of FeTi$_2$O$_5$.}
\end{center}
\end{figure}

\section{Discussion}
\subsection{Elastic anomalies in $C_{\Gamma}(T)$}

For $A$Ti$_2$O$_5$ ($A$ = Co, Fe), ultrasound velocity measurements revealed the Curie-type softening in $C_{55}(T)$ of CoTi$_2$O$_5$ above $T_N$, which ceases below $T_N$ [Fig. 3(b)], as well as the discontinuous anomalies at $T_N$ in $C_{\Gamma}(T)$ for $A$Ti$_2$O$_5$ [Figs. 2--5]. These elastic anomalies, observed above and at $T_N$, are likely of magnetic origin rather than being caused by oxygen cluster multipoles. Given the absence of orbital degeneracy at the $A^{2+}$ site [Fig. 1(*)], the spin degrees of freedom play a crucial role in these elastic behaviors. Such anomalies are attributed to magnetoelastic coupling acting on the exchange interactions, where the exchange striction arises from a modulation of the exchange interactions by ultrasound as follows: [\cite{Luthi}]

\begin{equation}
H_{exs} = \sum_{ij}[J({\bf \delta} + {\bf u}_i - {\bf u}_j) - J({\bf \delta})]{\bf S}_i {\bf \cdot} {\bf S}_j.
\label{eq:ES1}
\end{equation}
Here, $\delta = {\bf R}_i-{\bf R_j}$ is the distance between two magnetic ions, and ${\bf u}_i$ is the displacement vector for the ion with equilibrium position ${\bf R}_i$. When a sound wave with polarization ${\bf u}$ and propagation ${\bf k}$ is given by ${\bf u}$ = ${\bf u}_0$exp[i(${\bf k \cdot r}-\omega t$)], where ${\bf u}_0$ and $\omega$ are the respective amplitude and frequency, the exchange striction of Eq.~(\ref{eq:ES1}) is rewritten as [\cite{Stern}]

\begin{equation}
H_{exs} = \sum_{i}(\frac{\partial J}{\partial {\bf \delta}} {\bf \cdot u}_0)({\bf k \cdot \delta})({\bf S}_i {\bf \cdot S}_{i+\delta})e^{i({\bf k \cdot R}_i-\omega t)}.
\label{eq:ES2}
\end{equation}
Here, the exponential is expanded to first order because with a 30-MHz ultrasound frequency $k\delta\ll1$. Eq.~(\ref{eq:ES2}) indicates that both the longitudinal and transverse sound waves can couple to the spin system via the exchange striction mechanism, depending on the directions of polarization ${\bf u}$ and propagation ${\bf k}$ relative to the exchange path ${\bf \delta}$.

For $A$Ti$_2$O$_5$ with 1D $A^{2+}$ chains, one possible origin for the observed elastic anomalies is the spin-Peierls effect, where the lattice distortion along the spin chain direction occurs at the transition temperature $T_{sp}$. In the $T$ dependence of the elastic moduli of the spin-Peierls system, it is expected from the exchange striction mechanism [Eq.~(\ref{eq:ES2})] that only the compressive modulus along the spin chain direction has a pronounced anomaly at $T_{sp}$, and the shear moduli have no anomaly, as observed for the prototypical spin-Peierls compound CuGeO$_3$ [\cite{Saint-Paul,Poirier,Schwenk}]. Thus, for $A$Ti$_2$O$_5$, although a quasi-1D magnetic character has been experimentally suggested [\cite{Xu2}], the absence of an elastic anomaly at $T_N$ in the spin-chain-direction ($a$-axis) compressive modulus $C_{11}(T)$ [Figs. 2(a) and 4(a)] rules out a possible spin-Peierls transition at $T_N$. In addition, the Curie-type softening in the shear modulus $C_{55}(T)$ of CoTi$_2$O$_5$ above $T_N$ [Fig. 3(b)] should have an origin other than the spin-Peierls instability.

\subsection{Curie-type softening in $C_{\Gamma}(T)$}

In magnets, Curie-type softening in the $T$ dependence of the elastic modulus $C_{\Gamma}(T)$ emerges as a precursor to a structural transition, which is driven by the coupling of the lattice to the electronic degrees of freedom [\cite{Luthi,Kino,Kataoka,Hazama,Nii,Watanabe3,Watanabe4,Watanabe7,Watanabe8,Watanabe9,Kindler,Ramshaw}]. Therefore, the Curie-type softening observed in the symmetry-lowering elastic mode $C_{55}(T)$ of CoTi$_2$O$_5$ above $T_N$ should be a precursor to the symmetry-lowering lattice distortion at $T_N$. Such a precursor softening to the structural transition is well known to occur as a result of the Jahn--Teller effect in orbital-degenerate systems [\cite{Luthi}]. However, the absence of orbital degeneracy at the Co$^{2+}$ site in CoTi$_2$O$_5$ rules out such orbital effects [Fig. 1(*)]. Thus, the Curie-type softening in $C_{55}(T)$ of CoTi$_2$O$_5$ above $T_N$ is most likely driven by spin-lattice-coupled fluctuations, which are a precursor to a spin-driven magnetostructural transition at $T_N$. It is noted that the magnitude of the Curie-type softening in CoTi$_2$O$_5$ ($\Delta C_{55}/C_{55}\sim0.15$ $\%$) [Fig. 3(b)] is significantly smaller than that observed in the prototypical spin-JT system of chromite spinel $A$Cr$_2$O$_4$ ($\Delta C_{\Gamma}/C_{\Gamma}\sim50$ $\%$) [\cite{Watanabe3}]. For CoTi$_2$O$_5$, this relatively small softening above $T_N$ is consistent with the fact that the lattice distortion below $T_N$ is too small to have been resolved experimentally [\cite{Kirschner}].

For CoTi$_2$O$_5$, considering that the strain generated by ultrasound in $C_{55}$ ({\bf k} $\parallel$ {\bf a}, {\bf u} $\parallel$ {\bf c}) tilts the monoclinic angle $\beta$ between the $a$ and $c$ crystal axes [Fig. 1(b)], the Curie-type softening in $C_{55}(T)$ above $T_N$ should be a precursor to the $ac$-plane shear lattice distortion at $T_N$. It is important to note that, for CoTi$_2$O$_5$, the observed Curie-type softening in the $ac$-plane shear mode $C_{55}(T)$ above $T_N$ is inconsistent with the structural symmetry breaking caused by AF ordering at $T_N$ with propagation vector {\bf q} = $(\pm \frac{1}{2}, \frac{1}{2}, 0)$ [\cite{Kirschner}]. For the magnetostructural transition to the AF order with {\bf q} = $(\pm \frac{1}{2}, \frac{1}{2}, 0)$, it is expected that an $ab$-plane shear lattice distortion occurs at $T_N$, where the monoclinic angle $\gamma$ between the $a$ and $b$ crystal axes is tilted [Figs. 1(b)]. Therefore, for CoTi$_2$O$_5$, the observation of the Curie-type softening in $C_{55}(T)$ above $T_N$ suggests that the AF transition at $T_N$ coincides with a orthorhombic-to-triclinic lattice distortion, which tilts the angles $\beta$ and $\gamma$ [Fig. 1(b)].

In CoTi$_2$O$_5$, although the AF order with {\bf q} = $(\pm \frac{1}{2}, \frac{1}{2}, 0)$ is expected to coincide with the $ab$-plane shear lattice distortion at $T_N$, the $ab$-plane shear elastic modulus $C_{66}(T)$ shows no evidence of Curie-type softening above $T_N$ [Fig. 3(c)]. In the spin-lattice-coupled system undergoing a magnetostructural transition, the $T$ dependence of the elastic modulus $C_{\Gamma}(T)$ in the paramagnetic phase can be decomposed into a ``usual'' background component, $C^{(0)}_{\Gamma}(T)$, arising from lattice anharmonicity, and an ``unusual'' Curie component, $\delta C^C_{\Gamma}(T)$, originating from magnetostructural fluctuations: $C_{\Gamma}(T)=C^{(0)}_{\Gamma}(T)+\delta C^C_{\Gamma}(T)$. Here, $C^{(0)}_{\Gamma}(T)$ exhibits monotonic hardening with decreasing $T$ [\cite{Varshni}], while $\delta C^C_{\Gamma}(T)$ contributes to the Curie-type softening, following $\delta C^C_{\Gamma}(T)\sim-1/T$ [\cite{Watanabe3}]. In the case of CoTi$_2$O$_5$, the absence of observed Curie-type softening in $C_{66}(T)$ suggests that the contribution of the Curie component $\delta C^C_{66}(T)$, if present, is small and comparable to or smaller than that of the background $C^{(0)}_{66}(T)$.

For FeTi$_2$O$_5$, the absence of observed Curie-type softening in $C_{\Gamma}(T)$ [Figs. 4 and 5] suggests that the contribution of the Curie component $\delta C^C_{\Gamma}(T)$ is either negligible or smaller than that in CoTi$_2$O$_5$. However, previous experimental and theoretical studies have shown that its crystal and AF structures are identical to those of CoTi$_2$O$_5$ [\cite{Lang}], and earlier experimental evidence has verified the presence of magnetoelastic coupling [\cite{Xu}]. Thus, it is reasonable to expect a magnetostructural transition at $T_N$ in FeTi$_2$O$_5$, similar to that in CoTi$_2$O$_5$, although the spin-lattice coupling in FeTi$_2$O$_5$ is likely weaker than in CoTi$_2$O$_5$.

In the spin-JT system, the $T$ dependence of the elastic modulus $C_{\Gamma}(T)$ in the paramagnetic phase is explained by assuming a coupling of ultrasound with the magnetic ions through the exchange striction mechanism [Eq.~(\ref{eq:ES2})] [\cite{Luthi,Watanabe3,Watanabe7}]. Specifically, the Curie component $\delta C^C_{\Gamma}(T)$ in $C_{\Gamma}(T)$ of the spin-JT system is explained by assuming the coupling of ultrasound to the structural unit cells via the exchange striction mechanism, and the presence of exchange-striction-sensitive inter-unit-cell interactions. For the Curie-component-active elastic modulus $C_{\Gamma}(T)=C^{(0)}_{\Gamma}(T)+\delta C^C_{\Gamma}(T)$ in the prototypical spin-JT system $A$Cr$_2$O$_4$, the contribution of the Curie component $\delta C^C_{\Gamma}(T)$ is so large that the background component $C^{(0)}_{\Gamma}(T)$ can be assumed to be $T$ independent: $C^{(0)}_{\Gamma}(T)\simeq C^{(0)}_{\Gamma}(0)$ with $C^{(0)}_{\Gamma}(0)$ the background component at $T$ = 0 K [\cite{Watanabe3}]. In this case, the mean-field expression of $C_{\Gamma}(T)$ is written as
\begin{equation}
C_{\Gamma}(T) = C_{\Gamma}^{(0)}(0) \frac{T-T_c}{T-\theta}.
\label{eq:Curie1}
\end{equation}
Here, $\theta$ is the inter-unit-cell interaction, and $T_c=\theta + NG^2/C_{\Gamma}^{(0)}(0)$ is the second-order critical temperature for elastic softening $C_{\Gamma}\rightarrow$ 0 with $N$ the number density of the structural unit cell and $G$ the constant of coupling of the structural order parameter to the strain. $\theta$ is positive (negative) when the interaction is ferrodistortive (antiferrodistortive). Eq. (3) reproduces the observed Curie-type softening in $C_{\Gamma}(T)$ of $A$Cr$_2$O$_4$ very well [\cite{Watanabe3}].

\begin{figure}[t]
\begin{center}
\includegraphics[scale=0.7]{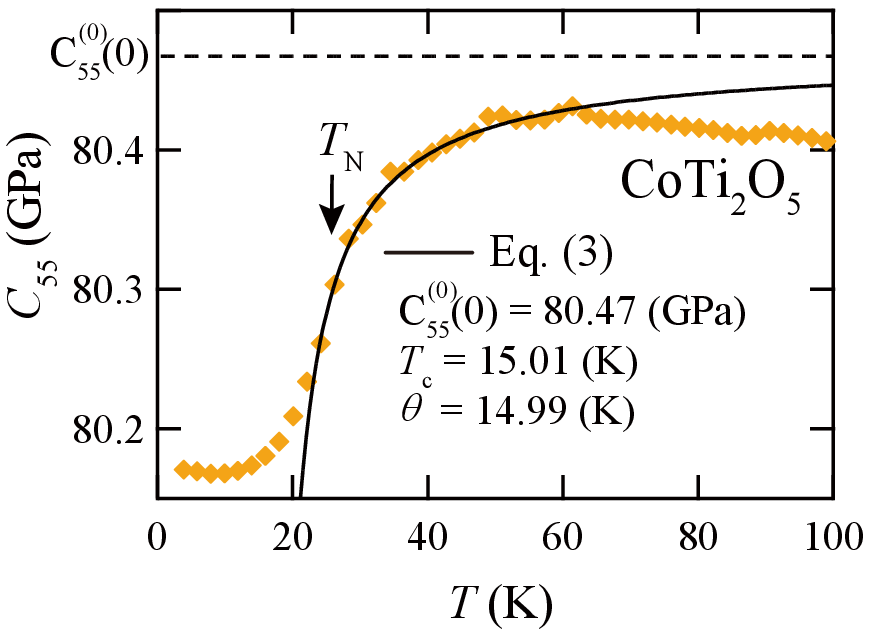}
\caption{\label{fig:fig6} (Color online) $C_{55}(T)$ below 100 K in CoTi$_2$O$_5$ [markers, from Fig. 3(b)]. The labeled arrows indicate $T_N \sim$ 26 K of CoTi$_2$O$_5$. The solid curve is a fit of Eq. (3) to the experimental $C_{55}(T)$ in the $T$ range $T_N \leq T \leq$ 60 K; the values for the fit parameters are also listed. The dashed horizontal line indicates $C_{55}^{(0)}(0)$ in Eq. (3).}
\end{center}
\end{figure}

In Fig. 6, a fit of Eq. (3) to the experimental $C_{55}(T)$ of CoTi$_2$O$_5$ in the $T$ range $T_N \leq T \leq$ 60 K is shown as a solid curve, with the corresponding fit parameters listed. The fit curve aligns well with the experimental data within this $T$ range. However, deviations from the fit become evident above $\sim$60 K, indicating that the $T$-dependent background $C^{(0)}_{55}(T)$ contributes to the experimental $C_{55}(T)$ in CoTi$_2$O$_5$.

For CoTi$_2$O$_5$, the small observed softening in $C_{55}(T)$ ($\Delta C_{55}/C_{55}\sim0.15$ $\%$), combined with the deviation of the experimental $C_{55}(T)$ from the Eq. (3) fit above $\sim$60 K [Fig. 6], suggests that the contribution of the Curie component $\delta C^C_{55}(T)$ is small and comparable to that of the background $C^{(0)}_{55}(T)$. Thus, for the Curie-component-active elastic modulus $C_{\Gamma}(T)=C^{(0)}_{\Gamma}(T)+\delta C^C_{\Gamma}(T)$ in $A$Ti$_2$O$_5$ ($A$ = Co, Fe), we should consider the $T$-dependent background $C^{(0)}_{\Gamma}(T)$. In this case, Eq. (3) is rewritten as
\begin{equation}
C_{\Gamma}(T) = C_{\Gamma}^{(0)}(T)-\frac{NG^2}{T-\theta},
\label{eq:Curie2}
\end{equation}
where $\delta C^C_{\Gamma}(T)=-\frac{NG^2}{T-\theta}$. For the experimental $C_{\Gamma}(T)$ in $A$Ti$_2$O$_5$, it is unfortunately not possible in the present study to determine the ``true'' $T$-dependent background $C_{\Gamma}^{(0)}(T)$. Consequently, the experimental $C_{\Gamma}(T)$ cannot be analyzed using Eq. (4). In future work, measurements of $C_{\Gamma}(T)$ in the nonmagnetic analogue of the orthorhombic pseudobrookite MgTi$_2$O$_5$ might help determine the background $C_{\Gamma}^{(0)}(T)$ for $C_{\Gamma}(T)$ in $A$Ti$_2$O$_5$ [\cite{Indo}], enabling a more accurate analysis of the experimental $C_{\Gamma}(T)$ for $A$Ti$_2$O$_5$ using Eq. (4).

\subsection{Non-Curie-type $C_{\Gamma}(T)$}

For $A$Ti$_2$O$_5$ ($A$ = Co, Fe), this study finds that only $C_{55}(T)$ of CoTi$_2$O$_5$ exhibits Curie-type softening above $T_N$, while other elastic moduli, $C_{\Gamma}(T)$, show hardening upon cooling [Figs. 2--5]. However, as discussed in Sec. IV B, the small magnitude of the observed softening in $C_{55}(T)$ of CoTi$_2$O$_5$ ($\Delta C_{55}/C_{55}\sim0.15$ $\%$) suggests that the contribution of the Curie component $\delta C^C_{55}(T)$ is small and comparable to that of the background $C^{(0)}_{55}(T)$. Therefore, it is necessary to investigate the potential presence of the Curie component $\delta C^C_{\Gamma}(T)$ in the ``non-Curie-type'' elastic moduli $C_{\Gamma}(T)$ of $A$Ti$_2$O$_5$, excluding the Curie-type $C_{55}(T)$ of CoTi$_2$O$_5$.

For the non-Curie-type elastic moduli $C_{\Gamma}(T)$ of $A$Ti$_2$O$_5$, we investigate the potential presence of the Curie component $\delta C^C_{\Gamma}(T)$ by fitting the experimental $C_{\Gamma}(T)$ to an empirical equation for the background $C^{(0)}_{\Gamma}(T)$ [\cite{Varshni}]:

\begin{equation}
C_{\Gamma}^{(0)}(T)=C_{\Gamma}^{(0)}(0)-\frac{A}{\exp(B/T)-1},
\label{eq:Curie1}
\end{equation}
where $C_{\Gamma}^{(0)}(0)$ is the background elastic modulus at $T$ = 0 K, and $A$ and $B$ are fitting parameters. In the case of a Curie-component-active elastic modulus, $C_{\Gamma}(T)=C_{\Gamma}^{(0)}(T)+\delta C^C_{\Gamma}(T)$, the contribution of the Curie component, $\delta C^C_{\Gamma}(T)\sim -1/T$, to $C_{\Gamma}(T)$ weakens at higher $T$'s, but strengthens at lower $T$'s. Consequently, fitting Eq. (5) to the Curie-component-active experimental $C_{\Gamma}(T)$ should yield good agreement at higher $T$'s, but deviations at lower $T$'s where $C_{\Gamma}^{(0)}(T)>C_{\Gamma}(T)$. Therefore, we fit Eq. (5) to the non-Curie-type experimental $C_{\Gamma}(T)$ of $A$Ti$_2$O$_5$ [Figs. 2--5, excluding Fig. 3(b)] in the higher $T$ range 200 K $\leq T \leq$ 300 K, assuming that $C_{\Gamma}^{(0)}(0)$ in Eq. (5) equals the value of the experimental $C_{\Gamma}(T)$ at the lowest measured temperature (2 K), $C_{\Gamma}^{2K}$, i.e., $C_{\Gamma}^{(0)}(0)=C_{\Gamma}^{2K}$.

In Fig. 7, the fits of Eq. (5) to the non-Curie-type experimental $C_{\Gamma}(T)$ of CoTi$_2$O$_5$, $C_{\Gamma}(T)$ excluding $C_{55}(T)$, in the $T$ range 200 K $\leq T \leq$ 300 K are shown as solid curves. The fitting parameter values are also listed in Fig. 7, where $C_{\Gamma}^{(0)}(0)$ in Eq. (5) is fixed at $C_{\Gamma}^{(0)}(0)=C_{\Gamma}^{2K}$. In Fig. 7, the fit curves for $C_{11}(T)$ [Fig. 7(a)(i)] and $C_{22}(T)$ [Fig. 7(a)(ii)] agree well with the experimental $C_{\Gamma}(T)$. However, the fit curves for $C_{33}(T)$ [Fig. 7(a)(iii)], $C_{44}(T)$ [Fig. 7(b)(i)], and $C_{66}(T)$ [Fig. 7(b)(ii)] begin to deviate from the experimental $C_{\Gamma}(T)$ below $\sim$ 200 K. These deviations are likely attributed to the presence of the Curie component $\delta C^C_{\Gamma}(T)$ or a more complex background $C_{\Gamma}^{(0)}(T)$ that is not fully captured by Eq. (5), possibly due to oxygen cluster multipoles.

In Fig. 8, the solid curves represent fits of Eq. (5) to the non-Curie-type experimental $C_{\Gamma}(T)$ of FeTi$_2$O$_5$ in the $T$ range 200 K $\leq T \leq$ 300 K. The fitting parameter values are also listed in Fig. 8, where $C_{\Gamma}^{(0)}(0)$ in Eq. (5) is fixed at $C_{\Gamma}^{(0)}(0)=C_{\Gamma}^{2K}$. The fit curves for $C_{11}(T)$ [Fig. 8(a)(i)] and $C_{22}(T)$ [Fig. 8(a)(ii)] agree well with the experimental $C_{\Gamma}(T)$, similar to the fits for $C_{11}(T)$ and $C_{22}(T)$ of CoTi$_2$O$_5$ [Figs. 7(a)(i) and 7(a)(ii), respectively]. Additionally, for FeTi$_2$O$_5$, the deviations in the fits for $C_{33}(T)$ [Fig. 8(a)(iii)] and $C_{44}(T)$ [Fig. 8(b)(i)] are notably smaller than those for CoTi$_2$O$_5$ [Figs. 7(a)(iii) and 7(b)(i)]. Among the elastic moduli $C_{11}(T)$--$C_{66}(T)$ of FeTi$_2$O$_5$ [Fig. 8], the fits for $C_{55}(T)$ [Fig. 8(b)(ii)] and $C_{66}(T)$ [Fig. 8(b)(iii)] exhibit noticeable deviations. These deviations are likely due to the presence of the Curie component $\delta C^C_{\Gamma}(T)$ or a more complex background $C_{\Gamma}^{(0)}(T)$ that is not fully described by Eq. (5), potentially caused by oxygen cluster multipoles. To address these deviations in FeTi$_2$O$_5$, as well as the previously noted deviations in $C_{33}(T)$ [Fig. 7(a)(iii)], $C_{44}(T)$ [Fig. 7(b)(i)], and $C_{66}(T)$ [Fig. 7(b)(ii)] of CoTi$_2$O$_5$, further studies are required. Future work should investigate whether these deviations originate from a dynamic elastic response or other factors, through analyses of the ultrasound frequency dependence of $C_{\Gamma}(T)$ and ultrasound attenuation.

\begin{figure}[t]
\begin{center}
\includegraphics[scale=0.6]{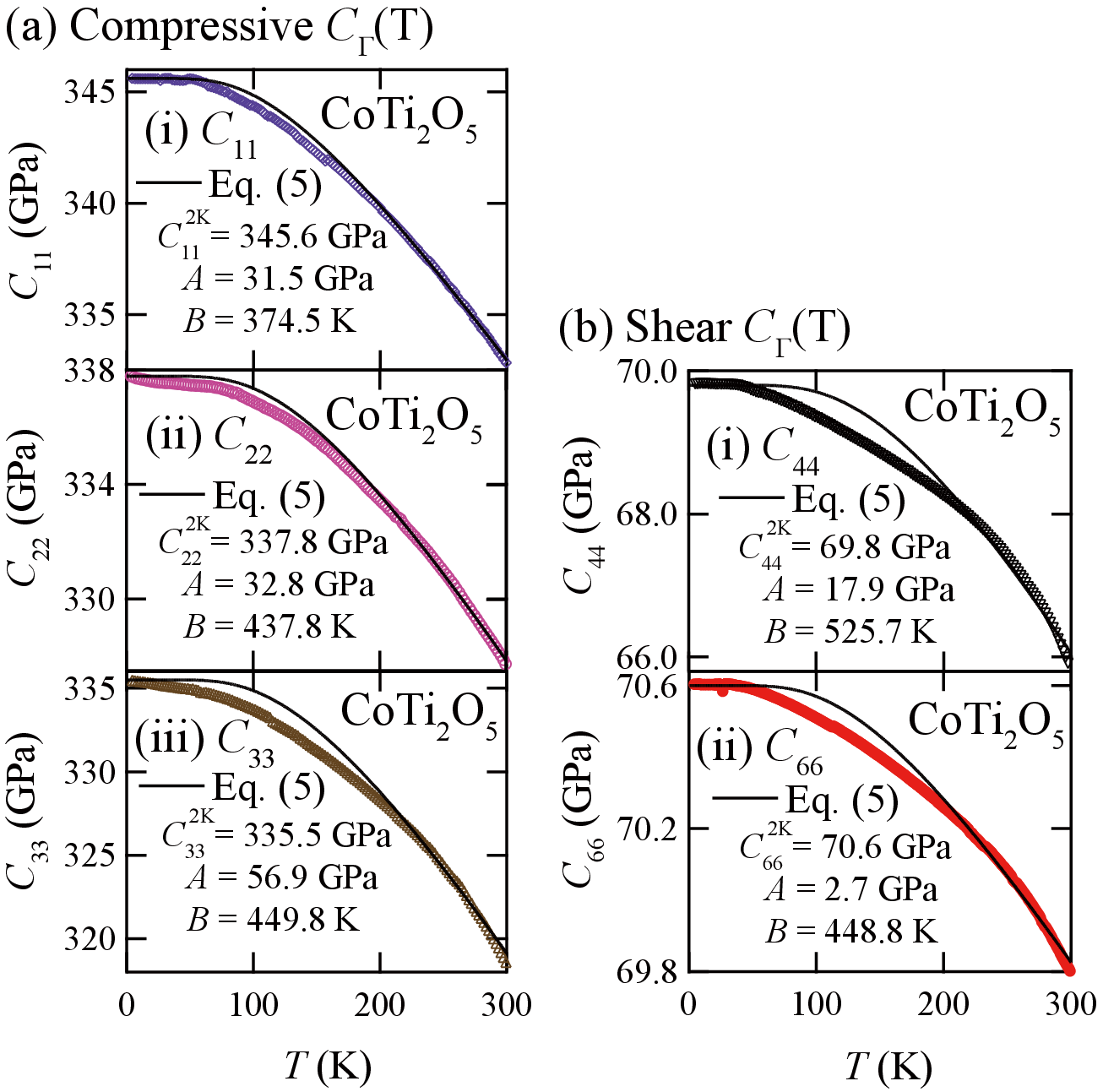}
\caption{\label{fig:fig6} (Color online) $C_{\Gamma}(T)$ of CoTi$_2$O$_5$: (a) compressive $C_{\Gamma}(T)$ of (i) $C_{11}(T)$, (ii) $C_{22}(T)$, and (iii) $C_{33}(T)$ [markers, from Fig. 2], and (b) shear $C_{\Gamma}(T)$ of (i) $C_{44}(T)$ and (ii) $C_{66}(T)$ [markers, from Fig. 3]. The solid curves are fits of Eq. (5) to the experimental $C_{\Gamma}(T)$ in the $T$ range 200 K $\leq T \leq$ 300 K with the assumption of $C_{\Gamma}^{(0)}(0)=C_{\Gamma}^{2K}$; the values for the fit parameters are also listed.}
\end{center}
\end{figure}

\begin{figure}[t]
\begin{center}
\includegraphics[scale=0.6]{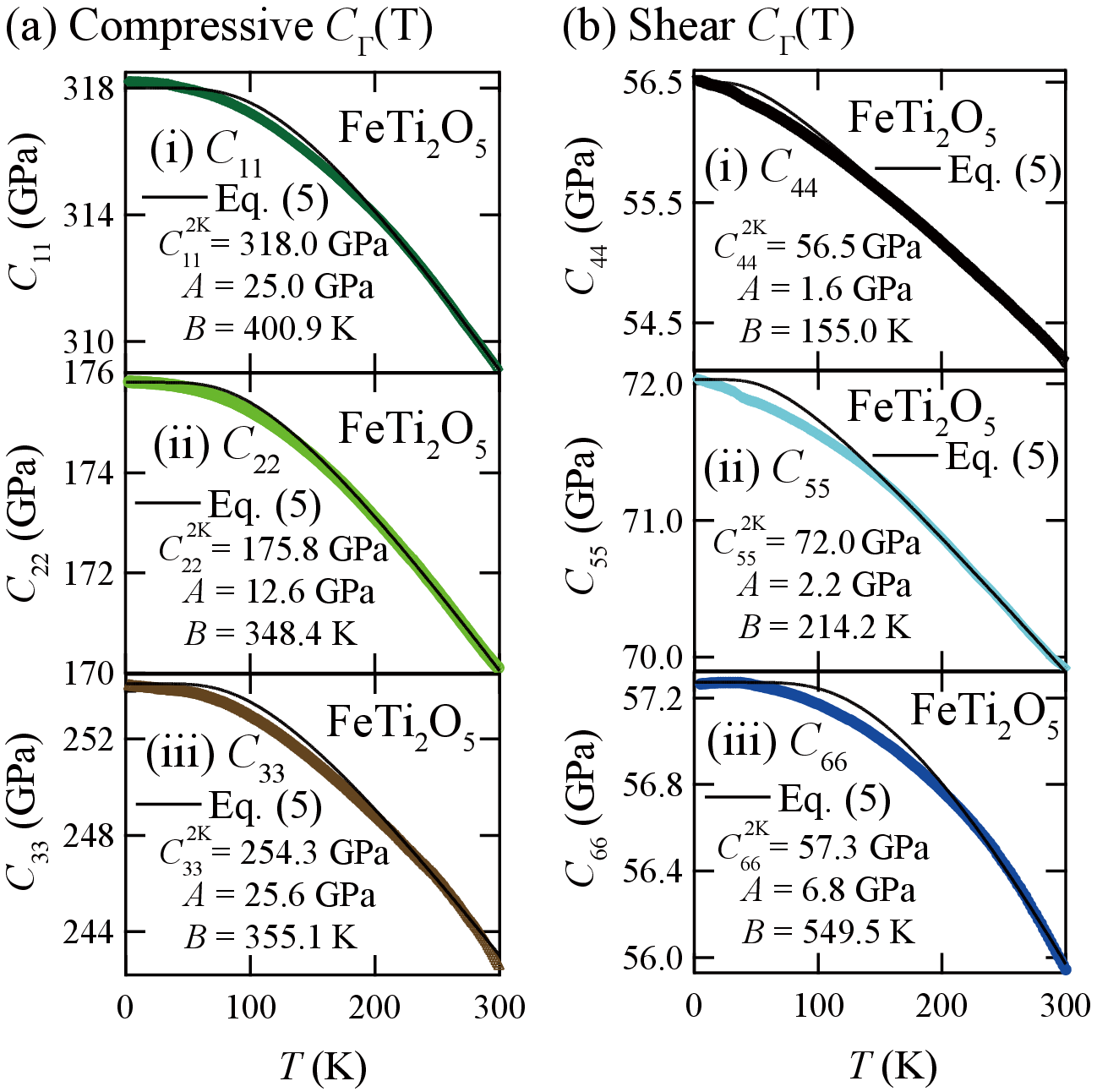}
\caption{\label{fig:fig7} (Color online) $C_{\Gamma}(T)$ of FeTi$_2$O$_5$: (a) compressive $C_{\Gamma}(T)$ of (i) $C_{11}(T)$, (ii) $C_{22}(T)$, and (iii) $C_{33}(T)$ [markers, from Fig. 4], and (b) shear $C_{\Gamma}(T)$ of (i) $C_{44}(T)$, (ii) $C_{55}(T)$, and (iii) $C_{66}(T)$ [markers, from Fig. 5]. The solid curves are fits of Eq. (5) to the experimental $C_{\Gamma}(T)$ in the $T$ range 200 K $\leq T \leq$ 300 K with the assumption of $C_{\Gamma}^{(0)}(0)=C_{\Gamma}^{2K}$; the values for the fit parameters are also listed.}
\end{center}
\end{figure}

\subsection{Magnetostructural fluctuations}

The present study reveals the presence of the Curie component, $\delta C^C_{\Gamma}(T)$, in $C_{55}(T)$ of CoTi$_2$O$_5$, where the Curie-type softening is directly observed [Fig. 3(b)]. Additionally, as discussed in Sec. IV C, the fit of Eq. (5) to the non-Curie-type experimental $C_{\Gamma}(T)$ shows noticeable deviations in $C_{33}(T)$, $C_{44}(T)$, and $C_{66}(T)$ for CoTi$_2$O$_5$ [Fig. 7], and in $C_{55}(T)$ and $C_{66}(T)$ for FeTi$_2$O$_5$ [Fig. 8]. This suggests the presence of either the Curie component $\delta C^C_{\Gamma}(T)$ or a more complex background $C_{\Gamma}^{(0)}(T)$ that Eq. (5) cannot fully capture.

For CoTi$_2$O$_5$, as discussed in Sec. IV B, the observed Curie-type softening in the $ac$-plane shear modulus $C_{55}(T)$ above $T_N$ [Fig. 3(b)] is inconsistent with the structural symmetry breaking caused by AF ordering at $T_N$ with {\bf q} = $(\pm \frac{1}{2}, \frac{1}{2}, 0)$ [\cite{Kirschner}]. This suggests that the lattice distortion at $T_N$ involves not only the $ac$-plane shear but also the $ab$-plane shear, indicating the orthorhombic-to-triclinic lattice distortion at $T_N$. Therefore, it is expected for CoTi$_2$O$_5$ that the Curie-component $\delta C^C_{\Gamma}(T)$ is present in not only the $ac$-plane shear $C_{55}(T)$ but also the $ab$-plane shear $C_{66}(T)$, indicating the presence of two distinct types of magnetostructural fluctuations in CoTi$_2$O$_5$ above $T_N$: the $ab$-plane shear fluctuations and the $ac$-plane shear fluctuations. The $ab$-plane fluctuations are likely a precursor to the AF order with {\bf q} = $(\pm \frac{1}{2}, \frac{1}{2}, 0)$, consistent with the spin-JT scenario [\cite{Kirschner}]. In contrast, the $ac$-plane fluctuations are newly identified in this study. Further experimental and theoretical investigations are needed to clarify the coexistence mechanism of the $ab$-plane and $ac$-plane shear magnetostructural fluctuations in CoTi$_2$O$_5$.

In CoTi$_2$O$_5$, according to the exchange striction mechanism described in Eq. (2), the transverse sound waves for $C_{55}$ ({\bf u} $\perp$ {\bf a}) and $C_{66}$ ({\bf k} $\perp$ {\bf a}) [Fig. 1] do not couple to the exchange interactions along the $a$ axis. Instead, they couple to the inter-$a$-axis-chain exchange interactions. Consequently, the Curie components in $C_{55}(T)$ and $C_{66}(T)$ of CoTi$_2$O$_5$ above $T_N$ likely originate from the coupling between the lattice and the inter-$a$-axis-chain exchange interactions. This suggests that, in CoTi$_2$O$_5$, while the exchange interactions are quasi-1D in nature, the magnetostructural transition at $T_N$ is driven by three-dimensional (3D) spin-lattice coupling involving the inter-$a$-axis-chain exchange interactions.

For FeTi$_2$O$_5$, the present study does not confirm the presence of magnetostructural fluctuations above $T_N$. However, previous experimental and theoretical studies have shown that its crystal and AF structures are identical to those of CoTi$_2$O$_5$ [\cite{Lang}], and earlier experimental evidence has verified the presence of magnetoelastic coupling [\cite{Xu}]. Therefore, it is reasonable to expect magnetostructural fluctuations in FeTi$_2$O$_5$ above $T_N$, similar to those in CoTi$_2$O$_5$, although the spin-lattice coupling in FeTi$_2$O$_5$ is likely weaker than in CoTi$_2$O$_5$. The difference in the strength of spin-lattice coupling between CoTi$_2$O$_5$ and FeTi$_2$O$_5$ is likely related to the distinct spin states of Co$^{2+}$ (3$d^7$, $S$ = 3/2) and Fe$^{2+}$ (3$d^6$, $S$ = 2) ions, respectively.

In the present study, the ultrasound velocity measurements were performed in zero magnetic field. Recent magnetostriction measurements on FeTi$_2$O$_5$ revealed that the intrachain $a$-axis and interchain $b$-axis exhibit negative and positive magnetostrictions, respectively, indicating the presence of magnetoelastic coupling [\cite{Xu}]. This finding suggests that, in $A$Ti$_2$O$_5$ ($A$ = Co, Fe), the magnetic field is likely to influence the spin-lattice-coupled frustration. Ultrasound velocity measurements on $A$Ti$_2$O$_5$ under magnetic field are anticipated to yield valuable insights into the effects of magnetic field on the frustration physics of these compounds.

\section{Summary}

Ultrasound velocity measurements of the orthorhombic pseudobrookites CoTi$_2$O$_5$ and FeTi$_2$O$_5$ revealed unusual elastic softness in CoTi$_2$O$_5$ above $T_N$ in the symmetry-lowering shear elastic mode, specifically in the $ac$-plane shear modulus $C_{55}$. For CoTi$_2$O$_5$, the observed elastic softness in $C_{55}$ above $T_N$ is inconsistent with the structural symmetry breaking caused by AF ordering at $T_N$. This suggests the presence of two distinct types of magnetostructural fluctuations above $T_N$, likely acting as a precursor to the symmetry-lowering lattice distortion at $T_N$. In contrast, the measurements for FeTi$_2$O$_5$ show that such elastic softness is either negligible or smaller, indicating weaker spin-lattice coupling. These findings highlight CoTi$_2$O$_5$ and FeTi$_2$O$_5$ as unique spin-lattice-coupled frustrated systems with low crystal symmetry, where, while the exchange interactions are quasi-1D in nature, the frustration is released by further lowering the crystal symmetry through 3D spin-lattice coupling, which is stronger in CoTi$_2$O$_5$ than in FeTi$_2$O$_5$.

\section{Acknowledgments}

This work was partly supported by a Grant-in-Aid for Scientific Research (C) (Grant No. 21K03476) from MEXT of Japan. DP acknowledges the Engineering and Physical Sciences Research Council (EPSRC), UK grant number EP/R024278/1 and the Oxford-ShanghaiTech collaboration project for financial support.

\end{document}